\documentclass{tlp}
\pdfoutput=1

\usepackage{url,alltt,amsmath,epsfig,epsf,amssymb,stmaryrd}

\usepackage{fancyvrb}

\SaveVerb{SVNId}|$Id: advancedICLP.tex 2052 2010-05-18 13:03:34Z frank $|

\usepackage[utf8]{inputenc}
\usepackage{enumerate}
\usepackage[all]{xy}
\usepackage{todo}
\usepackage{rotating}

\newtheorem{theorem}{Theorem}

\newtheorem{example}{Example}[section]
\newtheorem{definition}{Definition}[section]

\newcommand{\B}{\ensuremath{\mathbb{B}}}

\newcommand{\G}{\ensuremath{\mathbb{G}}}
\newcommand{\bL}{\ensuremath{\mathbb{L}}}
\newcommand{\mcP}{\ensuremath{\mathcal{P}}}
\renewcommand{\S}{\ensuremath{\mathbb{S}}}
\newcommand{\T}{\ensuremath{\mathbb{T}}}
\newcommand{\bP}{\ensuremath{\mathbb{P}}}
\newcommand{\V}{\ensuremath{\mathbb{V}}}
\newcommand{\CT}{\ensuremath{\mathcal{CT}}}

\newcommand{\bbB}{\ensuremath{\mathbb{B}}}

\newcommand{\bbT}{\ensuremath{\mathbb{T}}}
\newcommand{\bbV}{\ensuremath{\mathbb{V}}}

\newcommand{\subxt}{\left[X/t\right] }
\newcommand{\subXt}{\left[X/t\right] }

\newcommand{\Xet}{X\doteq t }

\newcommand{\bs}{{\bar s}}

\newcommand{\bang}{\ensuremath{!}}

\DeclareMathOperator{\chr}{chr}
\DeclareMathOperator{\id}{id}
\DeclareMathOperator{\vars}{vars}

\def\tuple#1{\langle #1 \rangle}
\newcommand{\stesq}[3]{\ensuremath{\tuple{#1; #2; #3}}}
\newcommand{\stbang}[4]{\ensuremath{\tuple{#1; #2; #3; #4}}}
\newcommand{\tstate}[6]{\ensuremath{\tuple{#1; #2; #3; #4}_{#5}^{#6}}}
\newcommand{\pstate}[6]{\ensuremath{\tuple{#1; #2; #3; #4}_{#5}^{#6}}}

\newcommand{\obang}{\ensuremath{{\omega_\bang}}}
\newcommand{\oesq}{\ensuremath{{\omega_e}}}

\newcommand{\Sbang}{\ensuremath{{\Sigma_{\bang}}}}
\newcommand{\Sesq}{\ensuremath{{\Sigma_e}}}

\newcommand{\ebang}{\ensuremath{\equiv_{\bang}}}
\newcommand{\eesq}{\ensuremath{\equiv_e}}

\newcommand{\der}{\ensuremath{\rightarrowtail}}

\newcommand{\derp}{\ensuremath{\der_p}}
\newcommand{\derbang}{\ensuremath{\der_{\bang}}}
\newcommand{\deresq}{\ensuremath{\der_e}}

\begin{document}
\title{A Complete and Terminating Execution Model for Constraint Handling Rules}

\author[Hariolf Betz, Frank Raiser and Thom Fr{\"u}hwirth]{HARIOLF BETZ,
FRANK RAISER \and THOM FR{\"U}HWIRTH\\
Faculty of Engineering and Computer Science, Ulm University, Germany\\
\email{firstname.lastname@uni-ulm.de}}

\maketitle

\textbf{Note: } This article has been published in \textit{Theory and Practice
of Logic Programming}, 10(4--6), 597--610, \copyright Cambridge University Press

\begin{abstract}
We observe that the various formulations of the operational semantics of
Constraint Handling Rules proposed over the years fall into a spectrum ranging
from the analytical to the pragmatic.
While existing analytical formulations facilitate program
analysis and formal proofs of program properties, they cannot be implemented as is.
We propose a novel operational semantics~$\obang$, which has a strong
analytical foundation, while featuring a terminating execution model.
We prove its soundness and completeness with respect to existing analytical
formulations and we provide an implementation in the form of a source-to-source
transformation to CHR with rule priorities.
\end{abstract}

\begin{keywords}
Constraint Handling Rules, Operational Semantics, Execution Model, Persistent
Constraints
\end{keywords}

\section{Introduction}
\label{sec:intro}
Constraint Handling Rules \cite{fruehwirth09} (CHR) is a declarative, multiset-
and rule-based programming language suitable for concurrent execution and
powerful program analysis. While it is known as a language that combines
efficiency with declarativity, publications in the field display a tendency to
favor one of these aspects over the other. We observe a spectrum of research
directions ranging from the \emph{analytical} to the \emph{pragmatic}.

On the analytical end of the spectrum, emphasis is put on CHR as a mathematical
formalism, declarativity, and the understanding of its logical foundations and
theoretical properties. Several formalizations of the operational semantics,
found in \cite{Fruhwirth1993,fruehwirth98} and \cite{fruehwirthabdennadher03},
belong to this side of the spectrum. Notable results building on these analytical
formalizations include decidable criteria for operational equivalence
\cite{abdennadherfruehwirth99} and confluence
\cite{abdennadherfruehwirthmeuss99}, a strong foundation of CHR in linear
logic~\cite{betzfruehwirth05}, as well as weak and strong parallelization, as
presented in \cite{fruehwirth05} and further developed toward concurrency in
\cite{Sulzmann2007,Sulzmann2008}.

A recent analytical formalization is the operational semantics~$\omega_e$, given
in \cite{Raiser2009a}. It consists in a rewriting system of equivalence classes
of states based on an axiomatic formulation of equivalence. It has been shown to
coincide with the operational semantics~$\omega_{va}$, which has been introduced
in \cite{fruehwirth09} to set a standard for all other operational semantics to
build upon.

On the downside, these operational semantics are detached from practical
implementation in that they are oblivious to questions of efficiency and
termination. Particularly, the class of rules called \emph{propagation rules}
causes trivial non-termination in both of them. Hence, it is safe to say that
the existing analytical formalizations of the operational semantics lack a 
terminating execution model.

This contrasts with most work on the pragmatic side of the spectrum, which
emphasizes practical implementation and efficiency over formal reasoning. It
originates with \cite{abdennadher97}, where a token-based approach is proposed in
order to avoid trivial non-termination: Every propagation rule is applicable only
once to a specific combination of constraints.  This is realized by keeping a
\emph{propagation history} -- sometimes called \emph{token store} -- in the CHR
state. Thus, we gain a terminating execution model for the full segment of CHR.

Building upon \cite{abdennadher97}, a plethora of operational semantics has been
brought forth, such as the token-based operational semantics~$\omega_t$ and its
refinement $\omega_r$ \cite{Duck2004}. The latter reduces non-determinism for a
gain in efficiency and sets the current standard for CHR implementations. Another
notable exponent is the priority-based operational semantics~$\omega_p$
\cite{dekoninckschrijversdemoen07}.

On the downside, token stores break with declarativity: Two states that differ
only in their token stores may exhibit different operational behavior while
sharing the same logical reading. Therefore, we consider token stores as
\emph{non-declarative elements} in CHR states.

Recent work on linear logical algorithms \cite{Simmons2008} and the close
relation of CHR to linear logic \cite{betzfruehwirth05} suggest a novel approach
that emphasizes aspects from both sides of the spectrum to a useful degree:
In this work, we introduce the notion of \emph{persistent constraints} to CHR, a
concept reminiscent of unrestricted or ``banged'' propositions in linear logic.
Persistent constraints provide a finite representation of the result of any
number of propagation rule firings.

We furthermore introduce a state transition system based on persistent
constraints, which is explicitly irreflexive. In combination, the two ideas
solve the problem of trivial non-termination while retaining declarativity and
preserving the potential for effective concurrent execution. This state
transition system requires no more than two rules. As every transition step
corresponds to a CHR rule application, it facilitates formal reasoning over
programs.

In this work, we show that the resulting operational semantics~$\obang$ is sound
and complete with respect to $\omega_e$. We show that $\obang$ can be faithfully
embedded into the operational semantics~$\omega_p$, thus effectively providing an
implementation in the form of a source-to-source transformation. All operational
semantics developed with an emphasis on pragmatic aspects lack this completeness
property. Therefore, this work is the first to show that it is possible to
implement CHR soundly and completely with respect to its abstract foundations,
whilst featuring a terminating execution model.

\begin{example}\label{ex:trans}
Consider the following straightforward CHR program for computing the transitive
hull of a graph represented by edge constraints~$e/2$: \[
\begin{array}{lclcl}
t & @ & e(X,Y), e(Y,Z) & \Longrightarrow & e(X,Z)
\end{array}
\]
This most intuitive formulation of a transitive hull is not a suitable
implementation in most existing operational semantics. In fact, for goals
containing cyclic graphs it is non-terminating in all aforementioned existing
semantics. In this work we show that execution in our proposed
semantics~$\obang$ correctly computes the transitive hull whilst guaranteeing
termination.
\end{example}

The remainder of this paper is structured as follows: We state the syntax of CHR
and summarize the existing operational semantics $\omega_t$ and $\omega_e$ in
Sect.~\ref{sec:opsems}. In Sect.~\ref{sec:ouromega}, we present our semantics
$\obang$, originally proposed in \cite{Betz2009}, and we state results concerning
its soundness and completeness with respect to $\oesq$. In
Sect.~\ref{sec:implementation}, we show how $\obang$ can be implemented by means
of a faithful source-to-source transformation into $\omega_p$. In
Sect.~\ref{sec:discussion}, we discuss the termination behavior of $\obang$ as
well as related work, before we conclude in Sect.~\ref{sec:conclusion}.
Proofs of the theorems presented in this work can be found in the accompanying
technical report \cite{Betz2010} \footnote{\cite{Betz2010} is available from
\url{http://vts.uni-ulm.de/doc.asp?id=7193}}, and will be omitted here.

\section{Preliminaries}
\label{sec:opsems}

We first introduce the syntax of CHR and the equivalence-based operational
semantics~$\oesq$, which offers a foundation for all other semantics, although it
lacks a terminating execution model. We furthermore present its refinements
$\omega_t$ and $\omega_p$.

\subsection{The Syntax of CHR}

Constraint Handling Rules distinguishes two kinds of constraints:
\emph{user-defined constraints} (or \emph{CHR constraints}) and \emph{built-in
constraints}. Reasoning on built-in constraints is possible through a
satisfaction-complete and decidable constraint theory~\CT.

CHR is a programming language that offers advanced rule-based multiset rewriting.
Its eponymous rules are of the form \[ r\ @\ H_1 \backslash H_2 \Leftrightarrow G
\mid B_c, B_b \] where $H_1$ and $H_2$ are multisets of user-defined constraints,
called the \emph{kept head} and \emph{removed head}, respectively. The
\emph{guard}~$G$ is a conjunction of built-in constraints and the
 \emph{body} consists of a conjunction of built-in constraints~$B_b$ and a
 multiset of
user-defined constraints~$B_c$. The \emph{rule name} $r$ is optional and may be
omitted along with the $@$ symbol. Note that throughout this paper, we omit the
curly braces around sets and multisets where there is no ambivalence. This
applies especially to CHR rules and states.

In this work, we put special emphasis on the class of rules where $H_2 =
\emptyset$, called \emph{propagation rules}. Propagation rules can be written
alternatively as \[r\ @\ H_1 \Longrightarrow G \mid B_c, B_b.\]

A \emph{variant} of a rule~$(r\ @\ H_1 \backslash H_2 \Leftrightarrow G \mid B_c,
B_b)$ with variables~$\bar x$ is a rule of the form $(r\ @\ H_1 \backslash H_2
\Leftrightarrow G \mid B_c,B_b)[\bar x / \bar y]$ for any sequence of pairwise
distinct variables~$\bar y$. For any rule~$(r\ @\ H_1 \backslash H_2
\Leftrightarrow G \mid B_c, B_b)$, the \emph{local variables}~$\bar l_r$ are
defined as $\bar l_r ::= \vars(G,B_c,B_b) \setminus \vars(H_1,H_2)$. A rule where
$\bar l_r = \emptyset$ is called \emph{range-restricted}.

A CHR program~$\mcP$ is a set of rules. A \emph{range-restricted} CHR program is
a set of range-restricted rules.

\subsection{Equivalence-based Operational Semantics $\oesq$}

In this section, we recall the \emph{equivalence-based operational
semantics~\oesq} \cite{Raiser2009a}. It is operationally close to the very
abstract semantics $\omega_{va}$, but we prefer it for its concise formulation
and the explicit distinction of global variables, user-defined, and built-in
constraints.

\begin{definition}[$\oesq$ State]\label{def:m_state}
An {\em \oesq\ state} is a tuple~$\stesq{\G}{\B}{\V}$. The \emph{user-defined
(constraint) store}~$\G$ is a multiset of CHR constraints. The {\em built-in
(constraint) store}~$\B$ is a conjunction of built-in constraints. $\V$ is a set
of variables called the \emph{global variables}. We use~$\Sesq$ to denote the set
of all $\oesq$ states. A variable~$v \in \B$ is called a \emph{strictly local
variable} iff $v \not \in (\V \cup \G)$.
\end{definition}

The operational semantics~\oesq\ is founded on equivalence classes of
states, based on the following definition of state equivalence.

\begin{definition}[$\oesq$ State Equivalence]
\label{def:m:equiv}
Equivalence between \oesq~states is the smallest equivalence relation~$\eesq$
over \oesq~states that satisfies the following conditions:

\begin{enumerate}
\item \label{cond:m:subst}
$\stesq{\G}{\Xet\wedge\B}{\V} \eesq \stesq{\G\subxt}{\Xet\wedge\B}{\V}$
\item \label{cond:m:ct} If
$\CT\models\exists \bar s.\B \leftrightarrow\exists\bar s'.\B'$ where $\bar s,
\bar s'$ are the strictly local variables of $\B,\B'$, respectively, then $
\stesq{\G}{\B}{\V} \eesq \stesq{\G}{\B'}{\V}$
\item \label{cond:m:global} 
If $X$ is a variable that does not occur in $\G$ or $\B$ then
$\stesq{\G}{\B}{\{X\}\cup\V} \eesq \stesq{\G}{\B}{\V}$
\item \label{cond:m:fail}
$\stesq{\G}{\bot}{\V} \eesq \stesq{\G'}{\bot}{\V}$
\end{enumerate}
\end{definition}

\begin{definition}[\oesq~Transitions]
	\label{def:opsem_classes}
For a CHR program~$\mcP$, the state transition system~$(\Sesq/\!\!\eesq,\deresq)$
is defined as follows. The transition is based on a variant of a rule~$r$ in
$\mcP$ such that its local variables are disjoint from the variables occurring in
the pre-transition state.

\begin{center}
\begin{tabular*}{9.5cm}{c}
$r\ @\ H_1 \setminus H_2 \Leftrightarrow G\mid B_c \uplus B_b$ \\[-.05cm]
\hline
$[\stesq{H_1 \uplus H_2 \uplus \G}{G\wedge\B}{\V}]
	\deresq^r
[\stesq{H_1 \uplus B_c \uplus \G}{G\wedge B_b\wedge\B}{\V}]$
\end{tabular*}
\end{center}

When the rule $r$ is clear from the context or not important, we may write
$\deresq$ rather than $\deresq^r$. By $\deresq^*$, we denote the
reflexive-transitive closure of $\deresq$.
\end{definition}

In the following, we freely mix equivalence classes and their representative,
i.e. we often write $\sigma \deresq \tau$ instead of $[\sigma] \deresq [\tau]$.

An inherent problem of $\oesq$ is its behavior with respect to propagation rules:
If a state can fire a propagation rule once, it can do so again and again, ad
infinitum. In the literature, this problem is referred to as \emph{trivial
non-termination} of propagation rules.

\begin{example}\label{ex:trans_omega_e}Reconsider the transitivity rule from
Example~\ref{ex:trans} and the following CHR state, which represents a cycle
consisting of two edges: \[ \sigma = \stesq{e(A,B),e(B,A)}{\top}{\emptyset}
\]
Let $t\ @\ e(A',B'), e(B',C') \Longrightarrow e(A',C')$ be a variant of the
transitivity rule, then it can be applied to $\sigma$, yielding an additional
loop edge:
\[
\begin{array}{ll}
\sigma \eesq & \stesq{e(A',B'), e(B',C')}{A=A' \land B=B' \land
A=C'}{\emptyset}\\
\der^t & \stesq{e(A',B'), e(B',C'), e(A',C')}{A=A' \land B=B'
\land A=C'}{\emptyset}\\
\eesq & \stesq{e(A,B), e(B,A), e(A,A)}{\top}{\emptyset}
\end{array}
\]
It is easily verified, that the transitivity rule can be applied again to
the same two constraints, yielding another~$e(A,A)$ constraint, hence this
program suffers from trivial non-termination in $\omega_e$.
\end{example}

\subsection{Operational Semantics with Rule Priorities}

The extension of CHR with rule priorities was initially proposed in
\cite{dekoninckschrijversdemoen07}. It annotates rules with priorities and
modifies the operational semantics such that among the applicable rules, we
always select one of highest priority for execution. The operational semantics of
this extension is denoted as $\omega_p$ and the formulation we use in work was
given in \cite{DeKoninck2008}.

The operational semantics~$\omega_p$ uses a so-called \emph{token store} to
avoid trivial non-termination. A propagation rule can only be applied once to each
combination of constraints matching the head. Hence, the token store keeps a
history of fired propagation rules based on constraint identifiers, as defined
below.

\begin{definition}[Identified CHR Constraints] An {\em identified CHR constraint}
$c\#i$ is a CHR constraint~$c$ associated with a unique integer~$i$, the {\em
constraint identifier}. We introduce the functions~$\chr(c\#i) = c$ and
$\id(c\#i) = i$, and extend them to sequences and sets of identified CHR
constraints in the obvious manner.
\end{definition}

The definition of an $\omega_p$~state is more complicated, because
identified constraints are distinguished from unidentified constraints and the
token store is added.

\begin{definition}[$\omega_p$~State] An {\em $\omega_p$~state} is a tuple of the
form~$\pstate{\G}{\S}{\B}{\T}{n}{\V}$ where the {\em goal (store)}~$\G$ is a
multiset of constraints, the {\em CHR (constraint) store}~$\S$ is a set of
identified CHR constraints, the {\em built-in (constraint) store}~$\B$ is a
conjunction of built-in constraints. The {\em token store} (or {\em propagation
history})~$\T$ is a set of tuples~$(r,I)$, where $r$ is the name of a propagation
rule and $I$ is an ordered sequence of constraint identifiers. $\V$ is a set of
variables called the \emph{global variables}. We use $\Sigma_p$ to denote the
set of all $\omega_p$~states.
\end{definition}

The corresponding transition system consists of the following three types of
transitions.

\begin{definition}[$\omega_p$~Transitions]
\label{def:omega-p-transitions}
For a CHR program~$\mcP$ with rule priorities, the state transition
system~$(\Sigma_p,\derp)$ is defined as follows.

\begin{description}
  \item[1. Solve.] $\tstate{\{c\} \uplus \G}{\S}{\B}{\T}{n}{\V} \derp
  \tstate{\G}{\S}{\B'}{\T}{n}{\V}$\\
  where $c$ is a built-in constraint and $\CT \models \forall((c \land \B)
  \leftrightarrow \B')$.
  \item[2. Introduce.] $\tstate{\{c\} \uplus \G}{\S}{\B}{\T}{n}{\V} \derp
  \tstate{\G}{\{c\#n\} \cup \S}{\B}{\T}{n+1}{\V}$\\
  where $c$ is a CHR constraint.
  \item[3. Apply.] $\tstate{\emptyset}{H_1 \cup H_2 \cup \S}{\B}{\T}{n}{\V}
  \derp \tstate{B}{H_1 \cup \S}{\Theta \land \B}{\T \cup t}{n}{\V}$ where $\mcP$
  contains a rule of priority~$p$ with fresh variables of the form \[p :: r\ @\
  H_1' \setminus H_2' \Leftrightarrow G \mid B\] and a matching
  substitution~$\Theta$ such that $\chr(H_1) = \Theta(H_1'), \chr(H_2) =
  \Theta(H_2')$, $\CT \models \exists (\B) \land \forall (\B \rightarrow
  \bar{\exists}_\B(\Theta \land G))$, $\Theta(p)$ is a ground arithmetic
  expression and $t = (r,\id(H_1)+\id(H_2)) \not \in \T$. Furthermore, no rule
  of priority~$p'$ and substitution~$\Theta'$ exists with $\Theta'(p') <
  \Theta(p)$ for which the above conditions hold.
\end{description} 
When the rule $r$ is clear from the context or not important, we may write
$\derp$ rather than $\derp^r$. By $\derp^*$, we denote the reflexive-transitive
closure of $\derp$.
\end{definition}

\section{Operational Semantics with Persistent Constraints $\obang$}
\label{sec:ouromega}

In this section, we present the operational semantics with persistent
constraints~\obang, proposed in \cite{Betz2009}. Our semantics is built on the
following basic ideas:

\begin{enumerate}
  \item In $\oesq$, the body of a
  propagation rule can be generated any number of times, provided that the corresponding head constraints are present in the
  store. In order to give consideration to this theoretical behavior, we
  introduce those body constraints as so-called \emph{persistent constraints}. A
  persistent constraint is a finite representation of a large, though unspecified
  number of identical constraints. For a proper distinction, constraints that are
  not persistent constraints are henceforth called \emph{linear} constraints.
  
  \item As a secondary consequence,
  arbitrary generation of rule bodies in $\oesq$ affects other types of CHR rules
  as well. Consider the following program:
  \[
	\begin{array}{lclcl}
	\text{r1} & @ & a & \Longrightarrow & b\\
	\text{r2} & @ & b & \Leftrightarrow & c
	\end{array}
  \]
  If executed with a goal $a$, this program can generate an arbitrary 
  number of constraints of the form $b$. As a consequence of this, it can
  also generate arbitrarily many constraints $c$. To take these indirect
  consequences of propagation rules into account, we introduce a rule's body
  constraints as persistent whenever its removed head can be matched completely
  with persistent constraints.
  
  \item As a persistent constraint represents an arbitrary number of identical
  constraints, we consider multiple occurrences of a persistent constraint as
  idempotent. Thus, we implicitly apply a set semantics to
  persistent constraints.
  
  \item We adapt the execution model such that a transition takes place only if
  the post-transition state is not equivalent to the pre-transition state. This
  entails two beneficial consequences: Firstly, in combination with the
  set semantics on persistent constraints, it avoids trivial non-termination of
  propagation rules. Secondly, as failed states are equivalent, it
  enforces termination upon failure.
\end{enumerate}

The formal definition of $\obang$ is given in Sect.~\ref{sec:definition}.
In Sect.~\ref{sec:sound-complete}, we state results concerning its soundness
and completeness with respect to $\oesq$.

\subsection{Definition}
\label{sec:definition}

In this section, we give a formal definition of our operational semantics
$\obang$. We present our adapted notions of state and state equivalence and a
transition system which consists of two distinct transition rules.

Definition~\ref{def:b_state} defines $\obang$ states. With respect to $\oesq$,
the goal store~$\G$ is split up into a store~$\bL$ of linear
constraints and a store~$\bP$ of persistent constraints:

\begin{definition}[$\obang$~State]\label{def:b_state} A \emph{$\obang$~state} is
a tuple of the form $\stbang{\bL}{\bP}{\B}{\V}$, where $\bL$ and $\bP$ are
multisets of CHR constraints called the \emph{linear (CHR) store} and
\emph{persistent (CHR) store}, respectively. $\B$ is a conjunction of built-in
constraints and $\V$ is a set of variables called the \emph{global variables}. We
use $\Sbang$ to denote the set of all \obang~states.
\end{definition}

Definition~\ref{def:b-vartypes} is analogous to $\oesq$, though adapted to
comply with Definition~\ref{def:b_state}.

\begin{definition}[Variable Types]\label{def:b-vartypes} For the variables
occurring in a \obang~state~$\sigma = \stbang{\bL}{\bP}{\B}{\V}$ we distinguish three different types:
\begin{enumerate}
  \item a variable~$v \in \V$ is called a {\em global} variable
  \item a variable~$v \not \in \V$ is called a {\em local} variable
  \item a variable~$v \not \in (\V \cup \bL \cup \bP)$ is called a {\em strictly
  local} variable
\end{enumerate}
\end{definition}

The following definition of state equivalence is adapted to comply with
Definition~\ref{def:b_state} and extended to handle idempotence of
persistent constraints.

\begin{definition}[Equivalence of $\obang$ States]
	\label{def:b:equiv}
Equivalence between $\obang$ states is the smallest equivalence
relation~$\ebang$ over $\obang$ states that satisfies the following
conditions:

\begin{enumerate}
\item \label{cond:b:subst} \emph{(Equality as Substitution)} Let $X$ be a
variable, $t$ be a term and $\doteq$ the syntactical equality relation.
\[
	\stbang{\bL}{\bP}{\Xet\wedge\B}{\V} \ebang
	\stbang{\bL\subXt}{\bP\subXt}{\Xet\wedge\B}{\V}
\]
\item \label{cond:b:ct} \emph{(Transformation of the Constraint Store)} If
$\CT\models\exists \bs.\B \leftrightarrow\exists\bar s'.\B'$ where $\bar s,
\bar s'$ are the strictly local variables of $\B,\B'$, respectively, then:
\[
	\stbang{\bL}{\bP}{\B}{\V} \ebang \stbang{\bL}{\bP}{\B'}{\V}
\]
\item \label{cond:b:global} \emph{(Omission of Non-Occurring Global Variables)}
If $X$ is a variable that does not occur in $\bL$, $\bP$, or $\B$ then:
\[
	\stbang{\bL}{\bP}{\B}{\{X\}\cup\V} \ebang \stbang{\bL}{\bP}{\B}{\V}
\]
\item \label{cond:b:fail} \emph{(Equivalence of Failed States)}
\[
	\stbang{\bL}{\bP}{\bot}{\V} \ebang \stbang{\bL'}{\bP'}{\bot}{\V'}
\]
\item \label{cond:b:cont} \emph{(Contraction)}
\[
	\stbang{\bL}{P\uplus P\uplus\bP}{\B}{\V} \ebang
	\stbang{\bL}{P\uplus\bP}{\B}{\V}
\]
\end{enumerate}
\end{definition}

Based on the definition of $\eesq$, we define the operational semantics~$\obang$
below. Since body constraints may be introduced either as linear or as persistent
constraints, uniform rule application is replaced by two distinct application
modes. Note that $\obang$ is only defined for \emph{range-restricted} programs.
In \cite{Betz2010} it is shown that $\obang$ is no longer compliant with
$\omega_e$ for non-range-restricted programs.

\begin{definition}[$\obang$ Transitions] \label{def:new_opsem}
For a range-restricted CHR program~$\mcP$, the state transition
system~$(\Sbang/\!\!\ebang,\derbang)$ is defined as follows.\\
\textbf{ApplyLinear:}
\begin{center}
\begin{tabular*}{9cm}{c}
$r\ @\ (H_1^l \uplus H_1^p) \backslash (H_2^l \uplus H_2^p) \Leftrightarrow G
\mid B_c, B_b \quad H_2^l \ne \emptyset \quad \sigma \ne \tau$\\[-.05cm]
\hline
$\sigma = [\stbang{H_1^l \uplus H_2^l \uplus \bL}{H_1^p \uplus H_2^p \uplus
\bP}{G \land \B}{\V}]$\\
$\derbang^r [\stbang{H_1^l \uplus B_c \uplus \bL}{H_1^p \uplus H_2^p \uplus
\bP}{G \land \B \land B_b}{\V}] = \tau$
\end{tabular*}
\end{center}
\textbf{ApplyPersistent:}
\begin{center}
\begin{tabular*}{8.25cm}{c}
$r\ @\ (H_1^l \uplus H_1^p) \backslash H_2^p \Leftrightarrow G \mid B_c, B_b
\quad \sigma \ne \tau$\\[-.05cm]
\hline
$\sigma = [\stbang{H_1^l \uplus \bL}{H_1^p \uplus H_2^p \uplus \bP}{G \land
\B}{\V}]$\\
$\derbang^r [\stbang{H_1^l \uplus \bL}{H_1^p \uplus H_2^p \uplus B_c \uplus
\bP}{G \land \B \land B_b}{\V}] = \tau$
\end{tabular*}
\end{center}

When the rule $r$ is clear from the context or not important, we may write
$\derbang$ rather than $\derbang^r$. By $\derbang^*$, we denote the
reflexive-transitive closure of $\derbang$.
\end{definition}

\begin{example}\label{ex:trans_obang}
Again consider the transitive edge program from
Example~\ref{ex:trans} and an analogous computation to that given in
Example~\ref{ex:trans_omega_e}, using an \textbf{ApplyPersistent} transition:
\[
\begin{array}{ll}
\sigma \ebang & \stbang{e(A',B'), e(B',C')}{\emptyset}{A=A' \land B=B'
\land A=C'}{\emptyset}\\
\der^t & \stbang{e(A',B'), e(B',C')}{e(A',C')}{A=A' \land B=B'
\land A=C'}{\emptyset}\\
\ebang & \stbang{e(A,B), e(B,A)}{e(A,A)}{\top}{\emptyset} = \sigma'
\end{array}
\]
The operational semantics~$\obang$ solves the trivial non-termination problem
through the combination of persistent constraints and its irreflexive transition
system, as the following observation shows:
\[
\begin{array}{ll}
\sigma' \ebang & \stbang{e(A',B'), e(B',C')}{e(A,A)}{A=A' \land B=B'
\land A=C'}{\emptyset}\\
\not \der^t & \stbang{e(A',B'), e(B',C')}{e(A,A), e(A',C')}{A=A'
\land B=B' \land A=C'}{\emptyset}\\
\ebang & \stbang{e(A,B), e(B,A)}{e(A,A)}{\top}{\emptyset} = \sigma'
\end{array}
\]
\end{example}

\subsection{Soundness and Completeness}
	\label{sec:sound-complete}

The following two theorems state the soundness and completeness of $\obang$
with respect to $\oesq$.

Theorem~\ref{thm:soundness} states that for every given state that can be
derived in $\obang$, we can derive a corresponding state in $\oesq$ which 
contains the linear constraints of the former state in equal multiplicities,
but its persistent constraints in arbitrarily high multiplicities.

\begin{theorem}[Soundness]\label{thm:soundness}Let
$\stbang{\G}{\emptyset}{\B}{\V},\stbang{\bL}{\bP}{\B'}{\V}\in\Sbang$. If
$\stbang{\G}{\emptyset}{\B}{\V}\derbang^* \stbang{\bL}{\bP}{\B'}{\V}$ then for
every $N\in\mathbb{N}$ there exists a state $\stesq{\G'}{\B'}{\V}\in\Sigma_e$
such that $\stesq{\G}{\B}{\V}\deresq^*\stesq{\G'}{\B'}{\V}$ and $\bL\uplus
N\cdot\bP\subseteq\G'$.
\end{theorem}

Theorem~\ref{thm:completeness} states that for every given state that can be
derived in $\oesq$, we can derive a corresponding state in $\obang$, such that 
its linear store and some subset of its persistent store add up exactly to the
user-defined store of the former state.

\begin{theorem}[Completeness]\label{thm:completeness}
Let $\stesq{\G}{\B}{\V},\stesq{\G'}{\B'}{\V}\in\Sesq$. If
$
	\stesq{\G}{\B}{\V}\deresq^* \stesq{\G'}{\B'}{\V}
$,
then there exists a state
$\stbang{\bL}{\bP}{\B'}{\V}\in\Sbang$ such that
$
	\stbang{\G}{\emptyset}{\B}{\V}\derbang^* \stbang{\bL}{\bP}{\B'}{\V}
$
and 
$
	\bL \subseteq \G' \subseteq \bP\uplus\bL
$.
\end{theorem}

\section{Implementation via Source-To-Source Transformation}
\label{sec:implementation}

In this section we provide an implementation of the operational
semantics~$\obang$ in the form of a source-to-source transformation. A CHR
program~$\mcP$ is transformed into a program~$\llbracket \mcP \rrbracket$ such
that $\llbracket \mcP \rrbracket$'s execution in $\omega_p$ is sound and complete
with respect to the execution of $\mcP$ in $\obang$.

The following definition of pathological rules is chosen such as to coincide with
those rules that cause redundant rule applications -- modulo state equivalence --
in $\oesq$, i.e. in a non-pathological program every rule applied to a
state~$\sigma$ results in a state~$\tau \not \eesq \sigma$ (cf.
\cite{Betz2010}). This ensures that \textbf{ApplyLinear}
transitions never fail due to irreflexivity, and hence, the resulting $\omega_p$
programs do not need to perform an explicit equivalence check.

\begin{definition}[Pathological Rules]\label{def:pathological}
A CHR rule $r\ @\ H_1 \backslash H_2 \Leftrightarrow G \mid B_c, B_b$ is
called \emph{pathological} if and only if $\exists \bbB .
\stesq{H_2}{\bbB \land G}{\emptyset} \eesq \stesq{B_c}{B_b}{\emptyset}$.
It is called \emph{trivially pathological} iff $\bbB=\top$. A CHR
program $\mcP$ is called pathological if it contains at least one pathological
rule.
\end{definition}

Assuming a CHR program~$\mcP$ without pathological rules, we now show how to
encode it as $\llbracket \mcP \rrbracket$ for execution in $\omega_p$.

For every $n$-ary constraint $c/n$ in $\mcP$, there exists a constraint~$c/(n+1)$
in $\llbracket \mcP \rrbracket$. In the following, for a multiset of user-defined
$\obang$-constraints~$M=\{c_1(\bar t_1), \ldots, c_n(\bar t_n)\}$ let $l(M) =
\{c_1(l, \bar t_1),\ldots,c_n(l, \bar t_n)\}$, $p(M) = \{c_1(p, \bar
t_1),\ldots,c_n(p, \bar t_n)\}$, and $c(M) = \{c_1(c, \bar t_1),\ldots,c_n(c,
\bar t_n)\}$.

The rules of $\llbracket \mcP \rrbracket$ are constructed via the following
source-to-source transformation.
\begin{enumerate}
  \item \label{b2r:applin} For every rule $r\ @\ H_1\setminus H_2
  \Leftrightarrow G\mid B$ in $\mcP$, and all multisets
  $H_1^l,H_1^p,H_2^l,H_2^p$ s.t. $H_1^l\uplus H_1^p = H_1$ and $H_2^l\uplus
  H_2^p = H_2$ and $H_2^l\neq\emptyset$, the following rule is in $\llbracket
  \mcP\rrbracket$:
  \[
  	3:: l(H_1^l)\uplus p(H_1^p)\uplus p(H_2^p)\setminus l(H_2^l) \Leftrightarrow
  	G\mid l(B_c), B_b
  \]
  \item \label{b2r:appper} For every rule $r\ @\ H_1\setminus H_2
  \Leftrightarrow G\mid B_c,B_b$ in $\mcP$, and all multisets $H_1^l,H_1^p$ s.t.
  $H_1^l\uplus H_1^p = H_1$, the following rule is in $\llbracket
  \mcP\rrbracket$:
  \[
  	3:: l(H_1^l)\uplus p(H_1^p)\uplus p(H_2) \Longrightarrow G\mid c(B_c),B_b
  \]
  \item \label{b2r:appcon} For every rule $\{c(p,\bar t),c(p,{\bar t}')\}\uplus H_1\setminus H_2
  \Leftrightarrow G\mid B$ in $\llbracket \mcP\rrbracket$, add also the
  following rule:
  \[
	3:: \{c(p,\bar t)\}\uplus H_1\setminus H_2 \Leftrightarrow \bar t={\bar
	t}'\wedge G\mid B
  \]
  \item \label{b2r:con} For every user-defined constraint $c/n$ in
  $\mcP$, add the following rules, where $\bar t$ is a sequence of $n$ different
  variables:
  \[
    \begin{array}{l}
	  1:: c(p, \bar t) \backslash c(c, \bar t) \Leftrightarrow \top\\
	  2:: c(c, \bar t) \Leftrightarrow c(p, \bar t)
	\end{array}
  \]
\end{enumerate}

\begin{example}[Encoding of Transitive Hull] We consider the transitive hull
program from Example~\ref{ex:trans}:
 \[
\begin{array}{lclcl}
t & @ & e(X,Y), e(Y,Z) & \Longrightarrow & e(X,Z)
\end{array}
\]
According to the encoding given above, the program is transformed as follows:
\[
\begin{array}{ccrcl}
3 & :: & e(l,X,Y), e(l,Y,Z) & \Longrightarrow & e(c,X,Z) \\
3 & :: & e(l,X,Y), e(p,Y,Z) & \Longrightarrow & e(c,X,Z) \\
3 & :: & e(p,X,Y), e(l,Y,Z) & \Longrightarrow & e(c,X,Z) \\
3 & :: & e(p,X,Y), e(p,Y,Z) & \Longrightarrow & e(c,X,Z) \\
\\
3 & :: & e(p,X,Y) & \Longrightarrow & X=Y \wedge Y=Z \mid e(c,X,Z) \\
\\
1 & :: & e(p,X,Y) \backslash e(c,X,Y) & \Leftrightarrow & \top\\
2 & :: & e(c,X,Y) & \Leftrightarrow & e(p,X,Y)
\end{array}
\]
The grouping of the rules above reflects the transformation
steps~\ref{b2r:appper}, \ref{b2r:appcon}, and~\ref{b2r:con}. Transformation
step~\ref{b2r:applin} is not productive in this example. The fifth rule above is
operationally equivalent to $3 :: e(p,X,X) \Longrightarrow e(c,X,X)$, and hence,
is redundant, as the resulting constraint will immediately be removed again by
the rule with priority~1. Furthermore, transformation step~\ref{b2r:appcon} also
adds an additional symmetric version of the fifth rule, which was omitted here,
as it is operationally equivalent as well.
\end{example}

Execution of a transformed program in $\omega_p$ is equivalent to execution
of the original program in $\obang$, as the following theorem shows.

\begin{theorem}[Soundness and Completeness of Encoding]
  Let $G,\bL,\bP$ be multisets of  user-defined constraints, $B, \bbB$
  conjunctions of built-in constraints, and $\bbV = \vars(G \land B)$. If $\mcP$
  is a non-pathologic CHR program, then \[
 \stbang{G}{\emptyset}{B}{\bbV} \derbang^*
  	\stbang{\bL}{\bP}{\bbB}{\bbV} \not \derbang \textrm{ in $\mcP$}
  \] \centerline{iff} \[
 \exists \bbT,n. \tstate{l(G), B}{\emptyset}{\top}{\emptyset}{0}{\bbV} \derp^*
 \tstate{\emptyset}{l(\bL)\uplus p(\bP)}{\bbB}{\bbT}{n}{\bbV} \not \derp \textrm{ in $\llbracket\mcP\rrbracket$}
   \]
\end{theorem}

\begin{example}[Example Runs of $\omega_p$ and $\obang$ Programs] The following
example derivation shows how the translated program terminates with a state that
corresponds with the result of an execution of the original program in $\obang$.
For clarity's and brevity's sake, we do not show all intermediate states and we
do not give the states' respective token stores explicitly.

\tiny\[
\begin{array}{ll}
& 	\tstate{e(l,A,B),e(l,B,A)}{\emptyset}{\top}{\emptyset}{0}{\{A,B\}}\\
\derp^* &
	\tstate{\emptyset}{e(l,A,B)\#0,e(l,B,A)\#1}{\top}{\emptyset}{2}{\{A,B\}}\\
\derp^{*} &
	\tstate{\emptyset}{e(l,A,B)\#0,e(l,B,A)\#1,e(c,A,A)\#2}{\top}{\ldots}{3}{\{A,B\}}\\
\derp^{*} &
	\tstate{\emptyset}{e(l,A,B)\#0,e(l,B,A)\#1,e(p,A,A)\#3}{\top}{\ldots)}{4}{\{A,B\}}\\
\derp^{*} &
	\tstate{\emptyset}{e(l,A,B)\#0,e(l,B,A)\#1,e(p,A,A)\#3,e(c,B,B)\#4}{\top}{\ldots}{5}{\{A,B\}}\\
\derp^{*} &
	\tstate{\emptyset}{e(l,A,B)\#0,e(l,B,A)\#1,e(p,A,A)\#3,e(p,B,B)\#5}{\top}{\ldots)}{6}{\{A,B\}}\\
\derp^{*} &
	\tstate{\emptyset}{e(l,A,B)\#0,e(l,B,A)\#1,e(p,A,A)\#3,e(p,B,B)\#5,e(c,A,B)\#6}{\top}{\ldots}{7}{\{A,B\}}\\
\derp^{*} &
	\tstate{\emptyset}{e(l,A,B)\#0,e(l,B,A)\#1,e(p,A,A)\#3,e(p,B,B)\#5,e(p,A,B)\#7}{\top}{\ldots)}{8}{\{A,B\}}\\
\derp^{*} &
	\tstate{\emptyset}{e(l,A,B)\#0,e(l,B,A)\#1,e(p,A,A)\#3,e(p,B,B)\#5,e(p,A,B)\#7,e(c,B,A)\#8}{\top}{\ldots}{9}{\{A,B\}}\\
\derp^{*} &
	\tstate{\emptyset}{e(l,A,B)\#0,e(l,B,A)\#1,e(p,A,A)\#3,e(p,B,B)\#5,e(p,A,B)\#7,e(p,B,A)\#9}{\top}{\ldots)}{10}{\{A,B\}}\\
\derp^{*} &
	\tstate{\emptyset}{e(l,A,B)\#0,e(l,B,A)\#1,e(p,A,A)\#3,e(p,B,B)\#5,e(p,A,B)\#7,e(p,B,A)\#9}{\top}{\ldots)}{24}{\{A,B\}}\\
\not\derp
\end{array}
\]
\normalsize
The above computation corresponds to the following execution
in $\obang$:
\[
\begin{array}{ll}
\sigma \ebang & \stbang{e(A,B), e(B,A)}{\emptyset}{\top}{\{A,B\}}\\
\derbang^t & \stbang{e(A,B),e(B,A)}{e(A,A)}{\top}{\{A,B\}}\\
\derbang^t & \stbang{e(A,B),e(B,A)}{e(A,A),e(B,B)}{\top}{\{A,B\}}\\
\derbang^t & \stbang{e(A,B),e(B,A)}{e(A,A),e(B,B),e(A,B)}{\top}{\{A,B\}}\\
\derbang^t &
\stbang{e(A,B),e(B,A)}{e(A,A),e(B,B),e(A,B),e(B,A)}{\top}{\{A,B\}}\\
\not \derbang
\end{array}
\] 
This example also demonstrates how $\obang$ streamlines execution which in turn
facilitates formal reasoning over derivations: the whole computation consists of
4 state transitions in $\obang$, whereas the corresponding computation in
$\omega_p$ requires 60 state transitions.
\end{example}

The presented source-to-source transformation satisfies conditions for an
\emph{acceptable encoding} according to \cite{Gabbrielli2009}, modulo the
necessary distinction between linear and persistent constraints in the
translation.

\section{Discussion}
\label{sec:discussion}

In this section, we discuss our insights on the behavior of $\obang$ in
comparison with existing operational semantics.

\subsection{Termination Behavior}\label{sec:termination}

Our proposed operational
semantics~$\obang$ exhibits a termination behavior different from $\omega_{t}$,
$\omega_p$, and $\oesq$. Compared to $\oesq$, we have solved the problem of
trivial non-termination of propagation rules, whereas any program terminating in
$\oesq$ also terminates in $\obang$. With respect to $\omega_t$ and $\omega_p$,
we found programs that terminate in $\obang$ but not in $\omega_t$ and
$\omega_p$, and vice versa.

We have seen in Example~\ref{ex:trans_omega_e} and Example~\ref{ex:trans_obang}
that the transitivity rule displays different behavior in $\oesq$ and $\obang$.
The program's termination behavior in $\omega_t$ and $\omega_p$ has been
investigated in \cite{Pilozzi2009}, where it is shown to terminate for acyclic
graphs. However, states containing cyclic graphs entail non-terminating behavior
(cf. \cite{Betz2010}). Contrarily, we show in the accompanying
technical report \cite{Betz2010} that in the operational semantics~$\obang$, 
the computation of the transitive hull terminates for every possible input.
At the same place, we present a CHR program that terminates in $\omega_t$ and
$\omega_p$, but not in $\obang$.

\subsection{Related Work}
\label{sec:related_work}

In \cite{Sarna-Starosta2007} the set-based semantics~$\omega_{set}$ has been
introduced. Its development was, among other considerations, driven by the
intention to eliminate the propagation history. Besides addressing the problem of
trivial non-termination in a novel manner, it reduces non-determinism similarly
to the refined operational semantics~$\omega_r$ \cite{Duck2004}. In
$\omega_{set}$, a propagation rule cannot be fired infinitely often for a
possible matching. However, multiple firings are possible, the exact number
depending on the built-in store. 

The authors of \cite{Sarna-Starosta2007} justify their set-based approach by the
following statement:
\begin{quotation} 
``When working with a multi-set-based constraint store, it appears that
propagation history is essential to provide a reasonable semantics.''
\end{quotation}
Our approach can be understood as a compromise since we avoid
a propagation history by imposing an implicit set semantics on persistent
constraints. The distinction between linear and persistent
constraints, however, allows us to restrict the set behavior to those
constraints, whereas the multiset semantics is
preserved for linear constraints.

Linear logical algorithms \cite{Simmons2008} (LLA) is a programming language
based on bottom-up reasoning in linear logic, inspired by logical algorithms
\cite{Ganzinger2002}. The first implementation of logical algorithms was realized
in CHR with rule priorities \cite{DeKoninck2009}. 

Our proposed operational semantics~$\obang$ is related to LLA \cite{Simmons2008},
but displays significant differences: Firstly, the notion of a constraint theory
with built-in constraints is absent in LLA. Secondly, LLA rules are restricted
such that persistent propositions cannot be derived multiple times, whereas
$\obang$ makes no such restriction and solves this problem via the irreflexive
transition system. Thirdly, LLA requires a strict separation of propositions into
linear and persistent ones. In $\obang$ a CHR constraint can occur in the linear
store, in the persistent store, or both.

On the other hand, the separation of
propositions in LLA allows the corresponding rules to freely mix linear and persistent
propositions in bodies. This is not directly possible with our approach, as CHR
constraints in a body are either added as linear or persistent constraints.

\section{Conclusion and Future Work}
\label{sec:conclusion}

The main motivation of this work was the observation that CHR research spans a
spectrum ranging from an analytical to a pragmatic end: on the analytical side of
the spectrum, emphasis is put on the formal aspects and properties of the
language while on the pragmatic side, it is put on implementation and efficiency.
A variety of operational semantics has been brought forth in the past, each
aligning with one side of the spectrum. In this work we proposed the novel
operational semantics~$\obang$, heeding both analytical and pragmatic aspects.

Unlike other operational semantics with a strong analytical foundation, $\obang$
thus provides a terminating execution model and may be implemented as is. We
provided evidence to this claim by presenting a sound and complete encoding of
$\obang$ into $\omega_p$, which can be used to implement $\obang$ by
source-to-source transformation.

Our operational semantics~$\obang$ is based on the concept of persistent
constraints. These are finite representations of an arbitrarily large number of
syntactically equivalent constraints. They enable us to subsume trivially
non-terminating computations in a single derivation step.

We proved soundness and completeness of our operational semantics~$\obang$ with
respect to $\omega_e$. The latter stands exemplarily for analytical
formalizations of the operational semantics, thus providing a strong analytical
foundation for $\obang$. This facilitates program analysis and formal proofs of
program properties.

In its current formulation, $\obang$ is only applicable to range-restricted CHR
programs -- a limitation we plan to address in the future. Furthermore, similar
to $\omega_t$ being the basis for numerous extensions to CHR
\cite{chr_survey_tplp08}, we plan to investigate the effect of building these
extensions on $\obang$.

In a concurrent environment, some kind of conflict resolution is required for the
case that multiple rules try to remove the same constraint. For example, in
\cite{Sulzmann2008} a transaction-based approach is used, leading to a rollback,
if the first evaluated rule application removed the constraint. The formulation
of the \textbf{ApplyPersistent} transition reveals that for persistent
constraints, no such conflicts have to be taken into account. A closer
investigation of potential benefits of the persistent constraint approach in
concurrent settings remains to be conducted.

\bibliography{bibliography}

\begin{thebibliography}{}

\bibitem[\protect\citeauthoryear{Abdennadher}{Abdennadher}{1997}]{abdennadher9%
7}
{\sc Abdennadher, S.} 1997.
\newblock Operational semantics and confluence of constraint propagation rules.
\newblock In {\em Principles and Practice of Constraint Programming}. 252--266.

\bibitem[\protect\citeauthoryear{Abdennadher and Fr{\"u}hwirth}{Abdennadher and
  Fr{\"u}hwirth}{1999}]{abdennadherfruehwirth99}
{\sc Abdennadher, S.} {\sc and} {\sc Fr{\"u}hwirth, T.} 1999.
\newblock Operational equivalence of {CHR} programs and constraints.
\newblock In {\em Principles and Practice of Constraint Programming, CP 1999},
  {J.~Jaffar}, Ed. Lecture Notes in Computer Science, vol. 1713.
  Springer-Verlag, 43--57.

\bibitem[\protect\citeauthoryear{Abdennadher, Fr{\"u}hwirth, and
  Meuss}{Abdennadher et~al\mbox{.}}{1999}]{abdennadherfruehwirthmeuss99}
{\sc Abdennadher, S.}, {\sc Fr{\"u}hwirth, T.}, {\sc and} {\sc Meuss, H.} 1999.
\newblock Confluence and semantics of constraint simplification rules.
\newblock {\em Constraints\/}~{\em 4,\/}~2, 133--165.

\bibitem[\protect\citeauthoryear{Betz and Fr{\"u}hwirth}{Betz and
  Fr{\"u}hwirth}{2005}]{betzfruehwirth05}
{\sc Betz, H.} {\sc and} {\sc Fr{\"u}hwirth, T.} 2005.
\newblock A linear-logic semantics for constraint handling rules.
\newblock In {\em Principles and Practice of Constraint Programming, 11th
  International Conference, CP 2005}, {P.~van Beek}, Ed. Lecture Notes in
  Computer Science, vol. 3709. Springer-Verlag, Sitges, Spain, 137--151.

\bibitem[\protect\citeauthoryear{Betz, Raiser, and Fr{\"u}hwirth}{Betz
  et~al\mbox{.}}{2009}]{Betz2009}
{\sc Betz, H.}, {\sc Raiser, F.}, {\sc and} {\sc Fr{\"u}hwirth, T.} 2009.
\newblock Persistent constraint in constraint handling rules.
\newblock In {\em Proceedings of 23rd Workshop on (Constraint) Logic
  Programming, WLP 2009}.
\newblock to appear.

\bibitem[\protect\citeauthoryear{Betz, Raiser, and Fr{\"u}hwirth}{Betz
  et~al\mbox{.}}{2010}]{Betz2010}
{\sc Betz, H.}, {\sc Raiser, F.}, {\sc and} {\sc Fr{\"u}hwirth, T.} 2010.
\newblock A complete and terminating execution model for constraint handling
  rules.
\newblock Tech. Rep.~01, Ulm University. January.

\bibitem[\protect\citeauthoryear{{De Koninck}}{{De
  Koninck}}{2009}]{DeKoninck2009}
{\sc {De Koninck}, L.} 2009.
\newblock Logical {A}lgorithms meets {CHR}: A meta-complexity result for
  {C}onstraint {H}andling {R}ules with rule priorities.
\newblock {\em Theory and Practice of Logic Programming\/}~{\em 9,\/}~2
  (March), 165--212.

\bibitem[\protect\citeauthoryear{{De Koninck}, Schrijvers, and Demoen}{{De
  Koninck} et~al\mbox{.}}{2007}]{dekoninckschrijversdemoen07}
{\sc {De Koninck}, L.}, {\sc Schrijvers, T.}, {\sc and} {\sc Demoen, B.} 2007.
\newblock User-definable rule priorities for {CHR}.
\newblock In {\em PPDP '07: Proceedings of the 9th ACM SIGPLAN international
  conference on Principles and practice of declarative programming}. ACM, New
  York, NY, USA, 25--36.

\bibitem[\protect\citeauthoryear{{De Koninck}, Stuckey, and Duck}{{De Koninck}
  et~al\mbox{.}}{2008}]{DeKoninck2008}
{\sc {De Koninck}, L.}, {\sc Stuckey, P.~J.}, {\sc and} {\sc Duck, G.~J.} 2008.
\newblock Optimizing compilation of {CHR} with rule priorities.
\newblock In {\em Functional and Logic Programming, 9th International Symposium
  (FLOPS)}, {J.~Garrigue} {and} {M.~V. Hermenegildo}, Eds. Lecture Notes in
  Computer Science, vol. 4989. Springer-Verlag, 32--47.

\bibitem[\protect\citeauthoryear{Duck, Stuckey, {Garc{\'i}a de la Banda}, and
  Holzbaur}{Duck et~al\mbox{.}}{2004}]{Duck2004}
{\sc Duck, G.~J.}, {\sc Stuckey, P.~J.}, {\sc {Garc{\'i}a de la Banda}, M.},
  {\sc and} {\sc Holzbaur, C.} 2004.
\newblock The refined operational semantics of {C}onstraint {H}andling {R}ules.
\newblock In {\em Logic Programming, 20th International Conference, ICLP 2004},
  {B.~Demoen} {and} {V.~Lifschitz}, Eds. Lecture Notes in Computer Science,
  vol. 3132. Springer-Verlag, Saint-Malo, France, 90--104.

\bibitem[\protect\citeauthoryear{Fr{\"u}hwirth}{Fr{\"u}hwirth}{1998}]{fruehwir%
th98}
{\sc Fr{\"u}hwirth, T.} 1998.
\newblock Theory and practice of constraint handling rules.
\newblock {\em Journal of Logic Programming, Special Issue on Constraint Logic
  Programming\/}~{\em 37,\/}~1-3 (October), 95--138.

\bibitem[\protect\citeauthoryear{Fr{\"u}hwirth}{Fr{\"u}hwirth}{2005}]{fruehwir%
th05}
{\sc Fr{\"u}hwirth, T.} 2005.
\newblock Parallelizing union-find in constraint handling rules using
  confluence analysis.
\newblock In {\em Logic Programming, 21st International Conference, ICLP 2005},
  {M.~Gabbrielli} {and} {G.~Gupta}, Eds. Lecture Notes in Computer Science,
  vol. 3668. Springer-Verlag, Sitges, Spain, 113--127.

\bibitem[\protect\citeauthoryear{Fr{\"u}hwirth}{Fr{\"u}hwirth}{2009}]{fruehwir%
th09}
{\sc Fr{\"u}hwirth, T.} 2009.
\newblock {\em {C}onstraint {H}andling {R}ules}.
\newblock Cambridge University Press.

\bibitem[\protect\citeauthoryear{Fr{\"u}hwirth and Abdennadher}{Fr{\"u}hwirth
  and Abdennadher}{2003}]{fruehwirthabdennadher03}
{\sc Fr{\"u}hwirth, T.} {\sc and} {\sc Abdennadher, S.} 2003.
\newblock {\em {E}ssentials of {C}onstraint {P}rogramming}.
\newblock Springer-Verlag.

\bibitem[\protect\citeauthoryear{Fr{\"u}hwirth and Hanschke}{Fr{\"u}hwirth and
  Hanschke}{1993}]{Fruhwirth1993}
{\sc Fr{\"u}hwirth, T.} {\sc and} {\sc Hanschke, P.} 1993.
\newblock Terminological reasoning with {C}onstraint {H}andling {R}ules.
\newblock In {\em Principles and Practice of Constraint Programming}. MIT
  Press, 80--89.

\bibitem[\protect\citeauthoryear{Gabbrielli, Mauro, and Meo}{Gabbrielli
  et~al\mbox{.}}{2009}]{Gabbrielli2009}
{\sc Gabbrielli, M.}, {\sc Mauro, J.}, {\sc and} {\sc Meo, M.~C.} 2009.
\newblock On the expressive power of priorities in {CHR}.
\newblock In {\em Proceedings of the 11th International ACM SIGPLAN Conference
  on Principles and Practice of Declarative Programming}, {A.~Porto} {and}
  {F.~J. L{\'o}pez-Fraguas}, Eds. ACM, Coimbra, Portugal, 267--276.

\bibitem[\protect\citeauthoryear{Ganzinger and McAllester}{Ganzinger and
  McAllester}{2002}]{Ganzinger2002}
{\sc Ganzinger, H.} {\sc and} {\sc McAllester, D.~A.} 2002.
\newblock Logical algorithms.
\newblock In {\em Logic Programming, 18th International Conference, ICLP 2002},
  {P.~J. Stuckey}, Ed. Lecture Notes in Computer Science, vol. 2401.
  Springer-Verlag, 209--223.

\bibitem[\protect\citeauthoryear{Pilozzi and {De Schreye}}{Pilozzi and {De
  Schreye}}{2009}]{Pilozzi2009}
{\sc Pilozzi, P.} {\sc and} {\sc {De Schreye}, D.} 2009.
\newblock Proving termination by invariance relations.
\newblock In {\em 25th International Conference Logic Programming, ICLP},
  {P.~M. Hill} {and} {D.~S. Warren}, Eds. Lecture Notes in Computer Science,
  vol. 5649. Springer-Verlag, Pasadena, CA, USA, 499--503.

\bibitem[\protect\citeauthoryear{Raiser, Betz, and Fr{\"u}hwirth}{Raiser
  et~al\mbox{.}}{2009}]{Raiser2009a}
{\sc Raiser, F.}, {\sc Betz, H.}, {\sc and} {\sc Fr{\"u}hwirth, T.} 2009.
\newblock Equivalence of {CHR} states revisited.
\newblock In {\em 6th International Workshop on Constraint Handling Rules
  (CHR)}, {F.~Raiser} {and} {J.~Sneyers}, Eds. 34--48.

\bibitem[\protect\citeauthoryear{Sarna-Starosta and
  Ramakrishnan}{Sarna-Starosta and Ramakrishnan}{2007}]{Sarna-Starosta2007}
{\sc Sarna-Starosta, B.} {\sc and} {\sc Ramakrishnan, C.} 2007.
\newblock Compiling {C}onstraint {H}andling {R}ules for efficient tabled
  evaluation.
\newblock In {\em 9th Intl.\ Symp.\ Practical Aspects of Declarative Languages,
  PADL}, {M.~Hanus}, Ed. Lecture Notes in Computer Science, vol. 4354.
  Springer-Verlag, Nice, France, 170--184.

\bibitem[\protect\citeauthoryear{Simmons and Pfenning}{Simmons and
  Pfenning}{2008}]{Simmons2008}
{\sc Simmons, R.~J.} {\sc and} {\sc Pfenning, F.} 2008.
\newblock Linear logical algorithms.
\newblock In {\em Automata, Languages and Programming, 35th International
  Colloquium, ICALP 2008}, {L.~Aceto}, {I.~Damg{\aa}rd}, {L.~A. Goldberg},
  {M.~M. Halld{\'o}rsson}, {A.~Ing{\'o}lfsd{\'o}ttir}, {and} {I.~Walukiewicz},
  Eds. Lecture Notes in Computer Science, vol. 5126. Springer-Verlag, 336--347.

\bibitem[\protect\citeauthoryear{Sneyers, Van~Weert, Schrijvers, and
  De~Koninck}{Sneyers et~al\mbox{.}}{2010}]{chr_survey_tplp08}
{\sc Sneyers, J.}, {\sc Van~Weert, P.}, {\sc Schrijvers, T.}, {\sc and} {\sc
  De~Koninck, L.} 2010.
\newblock As time goes by: {C}onstraint {H}andling {R}ules -- {A} survey of
  {CHR} research between 1998 and 2007.
\newblock {\em Theory and Practice of Logic Programming\/}~{\em 10,\/}~1,
  1--47.

\bibitem[\protect\citeauthoryear{Sulzmann and Lam}{Sulzmann and
  Lam}{2007}]{Sulzmann2007}
{\sc Sulzmann, M.} {\sc and} {\sc Lam, E. S.~L.} 2007.
\newblock A concurrent constraint handling rules semantics and its
  implementation with software transactional memory.
\newblock In {\em Proceedings of the POPL 2007 Workshop on Declarative Aspects
  of Multicore Programming}, {N.~Glew} {and} {G.~E. Blelloch}, Eds. ACM,
  19--24.

\bibitem[\protect\citeauthoryear{Sulzmann and Lam}{Sulzmann and
  Lam}{2008}]{Sulzmann2008}
{\sc Sulzmann, M.} {\sc and} {\sc Lam, E. S.~L.} 2008.
\newblock Parallel execution of multi-set constraint rewrite rules.
\newblock In {\em Proceedings of the 10th International ACM SIGPLAN Conference
  on Principles and Practice of Declarative Programming (PPDP)}, {S.~Antoy}
  {and} {E.~Albert}, Eds. ACM, Valencia, Spain, 20--31.

\end{thebibliography}

\end{document}